\documentclass[11pt]{article} 
\usepackage{hyperref} 
\pdfoutput=1 
 
\begin{document} 
 
\title{Electrokinetically Driven Reversible Banding of Colloidal Particles Near the Wall} 
 
\author{ Necmettin Cevheri and Minami Yoda \\ 
\\\vspace{6pt} G. W. Woodruff School of Mechanical Engineering, \\ Georgia Institute of Technology, Atlanta, GA 30332-0405, USA } 
\maketitle 
\begin{abstract} 
This fluid dynamics video, presented at the 66th Annual Meeting of the American Physical Society Division of Fluid Dynamics (Pittsburgh, PA), shows how dielectric colloidal polystyrene (PS) particles with a negative surface charge in a dilute suspension flowing through a microchannel can be manipulated and assembled into streamwise structures by using a combination of shear and an applied negative electric field along the streamwise direction.  Interestingly, these structures are only observed within about 1 $\mu$m of the wall, which also has a negative surface charge, and are not present in the bulk flow.
\end{abstract} 
\section{Introduction} 
This video is a visualization of 490 nm diameter fluorescently labeled PS spheres suspended in a 1 mmol/L sodium tetraborate aqueous solution at a volume fraction of $O(10^{-3})$ flowing through a fused-silica microchannel (nominal cross-sectional dimensions of 30 $\mu$m $\times$ 350 $\mu$m).  The visualizations use evanescent-wave illumination generated by the total internal reflection of light at the fused silica-fluid interface, and the exponentially decaying intensity of the illumination ensures that only particles within about 0.5 $\mu$m of the channel wall are visible in these images.  

The first set of video clips shows the near-wall particle dynamics in Poiseuille flow at a pressure gradient $\Delta p/L =$ 0.4 Bar/m.  In this set, flow goes from left to right, the vertical dimension of the field of view is about 200 $\mu$m, and the particles ``blink’’ due to Brownian diffusion.  The next segment shows what happens to the particle dynamics when an electric field along the streamwise direction of $E = -9$ V/cm (so the anode is upstream, and the cathode is downstream, and the mobile counterions, which are cations, are driven in a direction opposite to the Poiseuille flow).  More particles are visible, suggesting that the negatively charged particles are attracted to the negatively charged wall.  The next three segments show particle dynamics for Poiseuille flow at the same $\Delta p/L$ with an electric field $E = -23$ V/cm, $-40$ V/cm and -$47 V/cm$.  At $E = -23$ V/cm, even more particles are visible, and the formation of ``chains’’ of perhaps $4-6$ particles can be observed.  At the two highest values of $E$, dense streamwise bands of particles are evident, with a cross-stream dimension of roughly 10 $\mu$m.  At the end of the $E = -47$ V/cm clip, the electric field is turned off, and the bands disappear within about 15 s.

The next set of video clips shows the bands at a reduced magnification at different streamwise locations at $E = -47$ V/cm, where $x$ is measured from the bend at the upstream end of the straight section of this microchannel.  Flow still goes from left to right, and the entire 350 $\mu$m cross-stream dimension of the channel is visualized.  This set shows that the bands occur over most of the entire 4 cm length of the channel; the bands appear to become more defined as $x$ increases.

The last set of video clips (with a vertical dimension of $130 \mu$m) shows that the bands are deflected by obstacles, here debris stuck on the channel wall.  As a final note, our previous studies (Cevheri \& Yoda (2013) ``Evanescent-wave particle velocimetry studies of combined electroosmotic and Poiseuille flow,’’ Proceedings of the 10th International Symposium on Particle Image Velocimetry (Delft, the Netherlands)) show that applying a positive electric field to the same dilute particle suspension driven by a Poiseuille flow repels the particles from the negatively charged wall.

\end{document}